\documentclass[aps,amsmath,amssymb,prc,twosides,twocolumn,floats,superscriptaddress]{revtex4-2}

\usepackage{graphicx} 
\usepackage[utf8]{inputenc}
\usepackage[english]{babel}
\usepackage[shortlabels]{enumitem}
\usepackage{bm}
\usepackage{float}
\usepackage{tabularx}
\usepackage{array}
\usepackage[dvipsnames]{xcolor}
\usepackage{hyperref}

\usepackage{multirow}
\usepackage{longtable}
\usepackage{dcolumn}
\newcolumntype{d}{D{.}{.}{-1}}
\usepackage{setspace}
\usepackage{threeparttable}
\usepackage{tabularx}
\usepackage[symbol*]{footmisc}
\DefineFNsymbolsTM{myfnsymbols}{
  \textasteriskcentered *
  \textdagger    \dagger
  \textdaggerdbl \ddagger
  \textsection   \mathsection
  \textbardbl    \|%
  \textparagraph \mathparagraph
}%
\setfnsymbol{myfnsymbols}

\usepackage{siunitx}
\newcommand{\nuc}[2]{\hbox{$^{#1}$#2}}

\usepackage{gensymb}
\usepackage{mathtools}

\begin{document}

\title{A detailed view at magnetic dipole strengths: The case of semi-magic \nuc{50}{Ti}}

\author{B. Kelly}
\affiliation{Department of Physics, Florida State University, Tallahassee, Florida 32306, USA}

\author{M. Spieker}
 \email{Corresponding author: mspieker@fsu.edu}
\affiliation{Department of Physics, Florida State University, Tallahassee, Florida 32306, USA}

\author{U. Friman-Gayer}
\affiliation{Department of Physics, Institute for Nuclear Physics, Technische Universit{\"a}t Darmstadt, 64289 Darmstadt, Germany}

 \author{L.T. Baby}
\affiliation{Department of Physics, Florida State University, Tallahassee, Florida 32306, USA}

\author{T. Beck}
\email[]{Present address: KU Leuven, Celestijnenlaan 200c, 3001 Leuven, Belgium}
\affiliation{Department of Physics, Institute for Nuclear Physics, Technische Universit{\"a}t Darmstadt, 64289 Darmstadt, Germany}

 \author{A.L. Conley}
\affiliation{Department of Physics, Florida State University, Tallahassee, Florida 32306, USA}

\author{S. W. Finch}
\affiliation{Department of Physics, Duke University, Durham, North Carolina 27708-0308, USA}
\affiliation{Triangle Universities Nuclear Laboratory, Durham, North Carolina 27708-0308, USA}

\author{J. Isaak}
\affiliation{Department of Physics, Institute for Nuclear Physics, Technische Universit{\"a}t Darmstadt, 64289 Darmstadt, Germany}

\author{Krishichayan}
\affiliation{Department of Physics, Duke University, Durham, North Carolina 27708-0308, USA}
\affiliation{Triangle Universities Nuclear Laboratory, Durham, North Carolina 27708-0308, USA}

\author{E. Litvinova}
\affiliation{Department of Physics, Western Michigan University, Kalamazoo, Michigan 49008, USA}

\author{H. Pai}
\email[]{Present address: Extreme Light Infrastructure Nuclear Physics (ELI-NP), ``Horia Hulubei'' National Institute for R\&D in Physics and Nuclear Engineering (IFIN-HH), 077125 Bucharest-Magurele, Romania}
\affiliation{Department of Physics, Institute for Nuclear Physics, Technische Universit{\"a}t Darmstadt, 64289 Darmstadt, Germany}

\author{N. Pietralla}
\affiliation{Department of Physics, Institute for Nuclear Physics, Technische Universit{\"a}t Darmstadt, 64289 Darmstadt, Germany}

\author{D. Savran}
\affiliation{GSI Helmholtzzentrum f{\"u}r Schwerionenforschung GmbH, 64291 Darmstadt, Germany}

\author{W. Tornow}
\affiliation{Department of Physics, Duke University, Durham, North Carolina 27708-0308, USA}
\affiliation{Triangle Universities Nuclear Laboratory, Durham, North Carolina 27708-0308, USA}

\author{N. Tsoneva}
\affiliation{Extreme Light Infrastructure Nuclear Physics (ELI-NP), ``Horia Hulubei'' National Institute for R\&D in Physics
and Nuclear Engineering (IFIN-HH), 077125 Bucharest-Magurele, Romania}

\author{A. Volya}
\affiliation{Department of Physics, Florida State University, Tallahassee, Florida 32306, USA}

\author{V. Werner}
\affiliation{Department of Physics, Institute for Nuclear Physics, Technische Universit{\"a}t Darmstadt, 64289 Darmstadt, Germany}

\date{\today}

\begin{abstract}

Magnetic dipole, $M1$, strengths were studied in semi-magic \nuc{50}{Ti} up to the neutron-separation threshold by combining data from $(d,p)$ one-neutron transfer, $(\gamma,\gamma')$ real-photon scattering, $(e,e')$ inelastic scattering at extreme backward angles, and $(p,p')$ at $E_p = 210$\,MeV and extreme forward angles. The combination of all probes provided unique access to the neutron spin-flip contribution and the possibility to evaluate its role in generating the spin-flip $M1$ strengths. The small contribution of the neutron $(1f_{7/2})^{-1}(1f_{5/2})^{+1}$ spin-flip transitions, which were probed with the $(d,p)$ reaction, to the overall strength in \nuc{50}{Ti} questions the standard picture for the microscopic origin of spin-flip strength in the $fp$ shell. For \nuc{50}{Ti}, this letter shows that $J^{\pi} = 1^+$ states with larger neutron $(1f_{7/2})^{-1}(1f_{5/2})^{+1}$ spectroscopic factors do not correspond to the ones with the largest $B(M1;0^+_1 \rightarrow 1^+_i)$ strengths.

\end{abstract}


\maketitle

{\it Introduction.} During the last decade, the reduced magnetic dipole, $B(M1)$, strength of doubly magic \nuc{48}{Ca} has been subject of renewed interest with {\it ab-initio}-type approaches weighing in on the question of the significant quenching of the $B(M1)$ strength relative to the single-particle estimate \cite{Hol14a, Ach24a, Miy24a, Bra25a}. Different than for Gamow-Teller (GT) transitions \cite{Eks14a, Gys19a}, two-body currents were shown to have a minor impact on the $B(M1)$ strength in \nuc{48}{Ca} and different origins of the quenching were suggested \cite{Ach24a, Bra25a}. Traditionally, configuration mixing and ground-state correlations were discussed as one of the sources \cite{Gro79a, Bro83a, Kam84a, Tak88a, Kam89a, Hol14a, Gys19a, Ach24a, Hey10a, Rus13a, Gam20a, Gam22a}. It is, however, expected that the $B(M1)$ strength is largely generated by neutron $(1f_{7/2})^{-1}(1f_{5/2})^{+1}$ spin-flip transitions in the $N=28$ isotones \cite{Deh84a, Tak88a, Ach24a}, with the isoscalar contribution being smaller than 5\,$\%$ \cite{Bir16a}. In this letter, we will challenge this standard picture by using a large suite of experimental data for the $N=28$ isotone \nuc{50}{Ti}. The data, partly coming from new real-photon scattering and neutron-transfer experiments, allowed us to directly probe the microscopic origins of the strengths. Due to the impact of the $M1$ strength on nucleosynthesis and stellar processes (see, {\it e.g.}, \cite{Lan04a, Mat15a, Mat21a} and references therein), understanding these microscopic origins is crucial for fields beyond nuclear-structure physics.

The magnetic dipole, $M1$, operator has orbital ($\ell$) and spin-flip ($\sigma$) contributions in the isoscalar ($\Delta T =0$) and isovector ($\Delta T =1$) channels \cite{Ric90a, Hey10a,Mat21a}. Because of the involved nucleon $g$ factors, the isoscalar part is expected to contribute in a minor way to spin-flip $M1$ transitions \cite{Hey10a, Bir16a, Fuj00a, Lan04a}. For the stable, even-even $N=28$ isotones, selective proton-scattering experiments established that the fragmentation of the spin-flip $M1$ strength increases significantly with increasing proton number, $Z$ \cite{Dja83a}. In \nuc{48}{Ca}, Djalali {\it et al.} observed that the spin-flip $M1$ strength was concentrated in one dominant fragment \cite{Dja83a}. For the other $N=28$ isotones, this observation already hinted at the important role protons play in fragmenting the strength but also in possibly generating additional proton $(1f_{7/2})^{-1}(1f_{5/2})^{+1}$ spin-flip strength. The work of Liu and Zamick pointed out the need to include excitations of nucleons from the $1f_{7/2}$ orbital to the other $fp$ orbitals above the $Z=N=28$ shell closures, as well as the role of associated two-particle-two-hole (2p-2h) excitations, i.e., of more complex excitations to correctly describe the $M1$ strength fragmentation \cite{Liu87a}. Similar comments were made in \cite{Kam84a, Kam89a}. In addition to the spin-flip $M1$ strength, a comparably strong orbital $M1$ excitation was predicted at lower energies in $fp$-shell nuclei \cite{Ric90a, Hey10a}. The $J^{\pi} = 1^+$ state carrying the orbital $M1$ strength should result from the coupling of the neutron and proton quadrupole phonons, i.e., $[2^+_{\nu} \otimes 2^+_{\pi}]_{1^+}$. In early shell-model work, it was predicted to originate from the reorientation of nucleons in the $1f_{7/2}$ orbitals coupling to $J^{\pi} = 2^+$, respectively \cite{Hey10a, Liu87a}. As for the scissors mode in heavier nuclei\,\cite{Ric90a, Hey10a}, the associated orbital $B(M1)$ strength should increase with increasing quadrupole deformation. This was verified for \nuc{46,48}{Ti} by comparing data from the mentioned proton scattering experiments and complementary electron scattering experiments \cite{Ric90a}, which in contrast to proton scattering probed both the orbital and spin-flip contributions \cite{Ric95a}. Up to this point, a measurement of $B(M1)$ strengths below an excitation energy of 8\,MeV, where such orbital strength might exist, was missing for \nuc{50}{Ti}.

In this letter, we report new experimental and theoretical results on the complete $M1$ strength of semi-magic \nuc{50}{Ti} up to $E_x = 11.3$\,MeV by adding data below 8\,MeV. From a comparison of new real-photon $(\gamma,\gamma')$ scattering and one-neutron $(d,p)$ transfer data to available proton $(p,p')$ and electron $(e,e')$ scattering data, we identified an excited $1^+$ state with an unexpectedly large orbital $M1$ component. Using the new $(d,p)$ data, we also show that $1^+$ states with larger neutron $(1f_{7/2})^{-1}(1f_{5/2})^{+1}$ spectroscopic factors do not correspond to the ones with the largest $B(M1)$ strengths. As will be discussed, this finding is in conflict with model expectations.

{\it Experimental details.} The $(\gamma,\gamma')$ experiments were performed at the Darmstadt High Intensity Photon Source (DHIPS) of the Technical University of Darmstadt in Germany \cite{Son11a} and at the High Intensity $\gamma$-ray Source (HI$\gamma$S) of the Triangle Universities Nuclear Laboratory (TUNL) in Durham, NC, USA \cite{Wel09a}. At DHIPS, the photon beam is created through Bremsstrahlung. Thus, a continuous photon spectrum up to the energy of the electron beam provided by the Superconducting Darmstadt Linear Accelerator (S-DALINAC) is observed. End-point energies of 7.5\,MeV and 9.7\,MeV were chosen enabling the identification of some inelastic transitions, of feeding contributions to lower-lying states, and reliable photon flux calibration. To discriminate between $J^{\pi} = 1^-$ and $1^+$ states, a complementary experiment was performed at HI$\gamma$S using linearly polarized and quasi-monoenergetic photon beams from laser-Compton backscattering allowing the measurement of parity quantum numbers with the $\gamma^3$ setup\,\cite{loe13a}. For both experiments, a TiO$_2$ powder target with a density of 4.24 g/cm$^3$ was available. As the target was only enriched to 67.62(30)\,$\%$ in \nuc{50}{Ti} with \nuc{48}{Ti} being the main isotopic contaminant [24.06(30)\,$\%$], experiments were also performed with a highly-enriched \nuc{48}{Ti} target [99.81(3)\,$\%$]. In this letter, the complementary data were primarily used to identify states of \nuc{48}{Ti}, that were possibly populated in the $(\gamma,\gamma')$ experiments on the target which was only moderately enriched in \nuc{50}{Ti}, and for a consistency check to $\nuc{48}{Ti}(\gamma,\gamma')$ data reported previously by Degener {\it et al.}\,\cite{Deg90a}. Additional details can be found in \cite{Gay16a}. For more information on the nuclear resonance fluorescence (NRF) technique, see the review articles \cite{knei96a, zil22a}.

The $\nuc{49}{Ti}(d,p)\nuc{50}{Ti}$ one-neutron transfer experiment was performed at the Super-Enge Split-Pole Spectrograph (SE-SPS) of the John D. Fox Superconducting Linear Accelerator Laboratory at Florida State University (FSU)\,\cite{spi24a}. A 16-MeV deuteron beam was accelerated by the 9-MV Super-FN Tandem Van-de-Graaff accelerator and impinged on a self-supporting \nuc{49}{Ti} metal foil of areal density $\rho = 413$\,$\mu$g/cm$^2$. Protons were detected and their momenta analyzed by the position-sensitive focal-plane detector of the SE-SPS\,\cite{spi24a}. An average energy resolution of 50\,keV (FWHM) was achieved. Differential cross sections, $d\sigma/d\Omega$, were measured at laboratory scattering angles ranging from $15^{\circ}$ to $60^{\circ}$. Through a comparison to well-established reaction calculations with \textsc{fresco} \cite{fresco} using the adiabatic distorted wave approximation (ADWA) \cite{Wal76a} and global optical model potentials \cite{Kon03a}, the angular-momentum, $\ell$, transfers through which excited states of \nuc{50}{Ti} were populated in the $(d,p)$ reaction were determined. Spectroscopic factors, $S$, for neutron single-particle configurations were deduced by scaling the theoretically predicted to the experimentally measured angular distributions, i.e., $S = (d\sigma/d\Omega)_{exp}/(d\sigma/d\Omega)_{ADWA}$. More details of the $(d,p)$ experiment including all data will be presented in a forthcoming publication \cite{Kel25a}. Additional details on the determination of spectroscopic factors from these types of experiments performed at the FSU SE-SPS can be found in  \cite{Spi23a, spi24a}.

To identify $J^{\pi} = 1^-$ and $1^+$ states, which were excited with both probes, we used the following criteria: (a) the excitation energies must match within uncertainties, (b) the $\ell$ transfer observed in $(d,p)$ must allow for a $J^{\pi} = 1^-$ ($\ell =2$) or $1^+$ ($\ell =3$) assignment, (c) the $(\gamma,\gamma')$ angular distribution must establish $J=1$, and (d) the measurement of the analyzing power establishes the parity quantum number of the state. Criterion (d) is less strict as the parity can also be established by criterion (b) if criteria (a) and (c) hold. These criteria were successfully used in our previous studies combining $(d,p)$ and $(\gamma,\gamma')$ experiments \cite{Spi20a, Wei21a, Spi23a}. For $J^{\pi} = 1^+$ states, considered in this letter, we also used information from previous $(e,e')$ \cite{Sob85a} and $(p,p')$ \cite{Dja83a, Fuj85a} experiments.

\begin{figure}[t]
    \centering
    \includegraphics[width=0.99\linewidth]{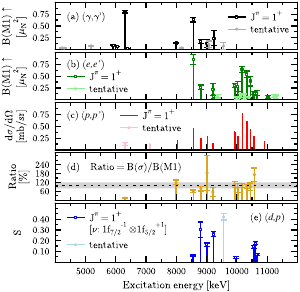}
    \caption{$B(M1;0^+_1 \rightarrow 1^+_i)$ strength distributions observed in (a) $(\gamma,\gamma')$ and (b) in the $(e,e')$ experiment of \cite{Sob85a}. (c) Differential cross sections, $d\sigma/d\Omega$, for $J^{\pi} = 1^+$ states populated in the $(p,p')$ experiment of \cite{Dja83a}. (d) Ratio $R$ of $d\sigma/d\Omega$ and $B(M1)\uparrow$ normalized to the respective values for the 10.17-MeV state. Gray band assumes a 20\,$\%$ uncertainty for the ratio. (e) Spectroscopic factors, $S$, for $J^{\pi} = 1^+$ states populated in the $(d,p)$ reaction via $\ell =3$ one-neutron transfer and assuming that the neutron got placed in the $1f_{5/2}$ orbital.}
    \label{fig:01}
\end{figure}

{\it Discussion and results.} The $(\gamma,\gamma')$, $(e,e')$ \cite{Sob85a}, $(p,p')$ \cite{Dja83a}, and $(d,p)$ data for $1^+$ states of \nuc{50}{Ti} are shown in Fig.\,\ref{fig:01}. First, we point out that we can exclude any significant feeding contributions to our newly determined $B(M1)$ strengths. As seen in Figs.\,\ref{fig:01}\,(a) and (b), agreement between the $B(M1)$ strengths determined in $(e,e')$ \cite{Sob85a} and our $(\gamma,\gamma')$ experiment is observed in the overlap region. This agreement also provides confidence in the $B(M1)$ strengths determined in the earlier $(e,e')$ experiment \cite{Sob85a}. A table with the $B(M1)$ strengths from $(\gamma,\gamma')$ can be found in the supplemental material \cite{suppl}.

The 6.307-MeV state is unambiguously identified as a $1^+$ state (see measured angular-distribution ratio and asymmetry in supplemental material \cite{suppl}). With $B(M1;0^+_1 \rightarrow 1^+_i) = 0.80(2)$\,$\mu^2_N$, it is one of the two most strongly populated states in the $(\gamma,\gamma')$ reaction. The other is the 8.563-MeV state with $B(M1) \uparrow \ = 0.63(5)$\,$\mu^2_N$. A $B(M1)\uparrow$ value of 0.85(11)\,$\mu_N^2$ was determined from the $(e,e')$ data\,\cite{Sob85a}. This is a difference of 25(11)\,$\%$. Within $2\sigma$, the two measurements agree with each other. Possibly, yet undetected $\gamma$-decay branching for the 8.563-MeV state could further improve the agreement. An increase of $\Gamma_0$ by only 8\,$\%$ is needed to obtain agreement within $1\sigma$. Data for the 6.307-MeV state were neither reported in \cite{Dja83a} nor in \cite{Sob85a}. Djalali {\it et al.} did, however, show a spectrum of their $\nuc{50}{Ti}(p,p')$ experiment, where the state appears to have been populated. We extracted a trace from a grayscale spectrum image of Fig.\,2 shown in \cite{Dja83a} and mapped pixel positions to the physical axes. This procedure allowed us to extract the spectrum from the figure, perform an energy calibration, and to determine the yield for the 6.307-MeV state relative to the 8.563-MeV state. As the differential cross section for the 8.563-MeV state was reported in \cite{Dja83a}, we could determine the differential cross section for the 6.307-MeV state from the relative yield, $d\sigma/d\Omega = 0.13(3)$\,mb/sr. The value was added to Fig.\,\ref{fig:01}\,(c). We cross-checked the validity of this approach with other states reported in \cite{Dja83a}.

The $B(M1)$ strength can be calculated from the orbital $B(l)$ and spin-flip $B(\sigma)$ parts \cite{Hey10a, Wil89a, Mat21a}. It has been shown that $(p,p')$ at the chosen kinematics primarily probes the spin-flip contribution, $B(\sigma)$, while $(\gamma,\gamma')$ and $(e,e')$ probe the entire $B(M1)$ strength \cite{Ric95a}. Fig.\,\ref{fig:01}\,(d) presents the ratio, $R$, of the differential cross sections $d\sigma/d\Omega$ from $(p,p')$ and the $B(M1)$ strengths from $(e,e')$, where both experimentally determined quantities were normalized to the respective values for the 10.17-MeV state before calculating $R$. The gray band in Fig.\,\ref{fig:01}\,(d) assumes a 20\,$\%$ uncertainty for the normalization, which is motivated by the comparably small deviation between the $(\gamma,\gamma')$ and $(e,e')$ results mentioned above. No uncertainties were reported for the $(p,p')$ data in \cite{Dja83a} but they are estimated at the same level from our spectrum analysis. The 10.17-MeV state lies in a region where spin-flip contributions are expected to dominate the $B(M1)$ strength. Large values of $R$ should, thus, be indicative of significant spin-flip contributions. As can be seen in Fig.\,\ref{fig:01}\,(d), almost all states above an energy of 8\,MeV have ratios consistent with being dominant spin-flip transitions, i.e., $R \geq 100$\,$\%$. The 6.307-MeV state with a ratio of $R=14(3)$\,$\%$ stands out. The ratio indicates significant orbital contributions to the $B(M1)$ value. For the states with large orbital contributions in \nuc{46}{Ti} and \nuc{48}{Ti}, the ratio $B(\sigma)/B(M1)$ is $\sim 12$\,$\%$ and $\sim 31$\,$\%$, respectively \cite{Ric90a}. Based on a comparison to \nuc{46,48}{Ti}, we estimate a significant $B(l)$ strength between 0.16 and 0.33\,$\mu_N^2$ for the 6.307-MeV state. For the 4.32-MeV, $J^{\pi} = 1^+$ state in \nuc{46}{Ti}, a $B(l)$ value of 0.43\,$\mu_N^2$ was reported \cite{Ric90a}. Qualitatively, the increase in orbital $M1$ collectivity agrees with the increase of quadrupole collectivity when going from \nuc{50}{Ti} to \nuc{46}{Ti} (see, {\it e.g.}, \cite{Gra24a}). Furthermore, and different from almost all of the other $1^+$ states, the 6.307-MeV state has appreciable $\gamma$-decay branches to the $J^{\pi} = 2^+_1$ and $J^{\pi} = 3^+_1$ states of 16.9(8)\,$\%$ and 6.5(12)\,$\%$, respectively, relative to the ground-state branch. The observations for the 6.307-MeV state clearly hint at a more complex microscopic structure than a simple 1p-1h neutron configuration.

\begin{figure*}
    \centering
    \includegraphics[width=1\linewidth]{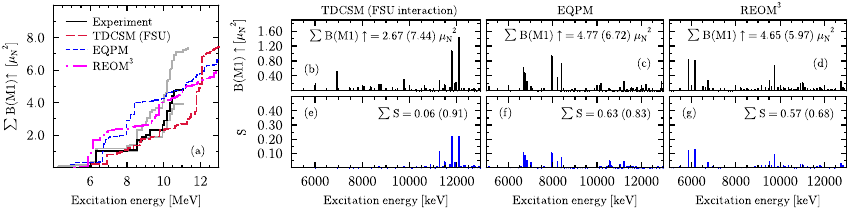}
    \caption{(a) running sum of $B(M1)$ strength predicted by the time-dependent continuum nuclear shell model (TDCSM) with the FSU interaction, the energy density functional theory plus quasiparticle phonon model approach (EQPM), and the relativistic equation of motion framework (REOM$^3$) up to an excitation energy of $E_x = 13$\,MeV. Experimental data are also shown. The gray, solid lines show the experimental lower and upper limit for the $B(M1)$ strength. (b)-(d) fragmentation of the total $B(M1)$ strength predicted by these models. The summed $B(M1)$ strengths up to $E_x = 11$\,MeV and 13\,MeV (in parenthesis) are given. (e)-(g) spectroscopic factors, $S$, for the overlap of the $\nuc{49}{Ti}(\mathrm{g.s.})+\nu:1f_{5/2}$ configuration with the $J^{\pi} =1^+$ wavefunction. The summed spectroscopic factors, $\sum S$, up to $E_x = 11$\,MeV and 13\,MeV (in parenthesis) are given.}
    \label{fig:02}
\end{figure*}

We now discuss the implications of our new $(d,p)$ data for the structure of $1^+$ states. As mentioned in the introduction, it might be expected that the $(1f_{7/2})^{-1}(1f_{5/2})^{+1}$ neutron 1p-1h, spin-flip component dominates the spin-flip strength. If we assume that \nuc{49}{Ti} with $J^{\pi}(g.s.) = 7/2^-$ has a $(\pi:1f_{7/2})^{2}_{J=0}(\nu:1f_{7/2})^7_{J=7/2}$ ground-state configuration, then this component could be directly accessed from the $(d,p)$ transfer data.  Recent Coulomb excitation experiments suggest that the stated configuration indeed dominates the \nuc{49}{Ti} ground-state wavefunction \cite{Gra24a}. In a forthcoming publication, we will show that we collected a total of $63(6)$\,$\%$ of the $1f_{5/2}$ neutron-adding strength for resolved excited states in \nuc{50}{Ti}\,\cite{Kel25a}. The centroid energy of this $\ell = 3$ strength is 8200(36)\,keV. The strength fraction is consistent with the quenching of the single-particle strengths in nuclei\,\cite{Kay13a, Kay22a}. Without quenching, summing over all spectroscopic factors for a given spin and parity quantum number should return a value of 1 or smaller \cite{satchler}. For the eight $1^+$ states shown in Fig.\,\ref{fig:01}\,(e), which were observed in $(d,p)$ and firmly identified with at least one of the other probes, a summed spectroscopic factor of $S = 1.2(2)$ is determined. This value appears consistent with unity. We are, thus, confident that we collected all $1f_{5/2}$ neutron-adding strength for $1^+$ states. The centroid for the associated $\ell =3$ strength is 9524(47)\,keV, i.e., significantly higher than the centroid of the entire $\ell =3$ strength. The $(d,p)$ angular distributions can be found in the supplemental material \cite{suppl}.

To better understand the influence of the neutron $(1f_{7/2})^{-1}(1f_{5/2})^{+1}$ 1p-1h, spin-flip component on the $B(M1)$ strength based on experimental data, a closer inspection of Fig.\,\ref{fig:01} is instructive. First, we see that the 6.307-MeV state is not populated in $(d,p)$. This means that this state does not have a $1f_{5/2}$ $(\ell =3)$ neutron particle component which can be reached from the \nuc{49}{Ti} ground state. The observation appears consistent with its more complex microscopic character mentioned above. The 8.563-MeV is only weakly populated in $(d,p)$ [see Fig.\,\ref{fig:01}\,(e)]. Most importantly, no state with a strong spectroscopic factor for placing a $1f_{5/2}$ neutron on top of the \nuc{49}{Ti} ground state, i.e., a neutron $(1f_{7/2})^{-1}(1f_{5/2})^{+1}$ 1p-1h structure, is observed in the $(d,p)$ reaction that at the same time carries significant $B(M1)$ strength.

We now look at details of the summed $B(M1)$ strengths to further quantify the contribution of the neutron $(1f_{7/2})^{-1}(1f_{5/2})^{+1}$ spin-flip component. For the firmly assigned $1^+$ states, the summed $B(M1)$ strength is 4.9(9)\,$\mu_N^2$. The centroid energy for that strength is $E_x = 9074(52)$\,keV. When tentative $1^+$ assignments are included, the summed $B(M1)$ strength increases to an upper limit of 7.4\,$\mu_N^2$. This includes three $J=1$ states with unknown parity below 5.5\,MeV. This summed strength appears consistent with $(\gamma,\gamma')$, $(e,e')$, and $(p,p')$ data reported for other $N=28$ isotones in \cite{Ste83a, Sob85a, Dja83a, Bir16a, Shi17a, Wil18a}. The eight states firmly identified as $1^+$ states in $(d,p)$ have a summed $B(M1)$ strength of 1.9(5)\,$\mu_N^2$. This corresponds to only $38^{+21}_{-19}$\,$\%$ of the total $B(M1)$ strength observed in \nuc{50}{Ti} up to $E_x = 11.3$\,MeV ($S_n = 10.9$\,MeV). Given that the $(1f_{7/2})^{-1}(1f_{5/2})^{+1}$ spectroscopic strength is only fragmented among a few states, it is surprising that they contribute so little to the $M1$ strength. For completeness, we note that the $(p,p')$ data of \cite{Dja83a} showed that no significant spin-flip $M1$ strength was missed up to $E_x = 15$\,MeV. Any remaining strengths must be fragmented to much higher energies and, at least for \nuc{50}{Ti}, not be of simple $(1f_{7/2})^{-1}(1f_{5/2})^{+1}$ neutron spin-flip character as the sum rule for that strength is exhausted.

\begin{figure}[t]
    \includegraphics[width=1\linewidth]{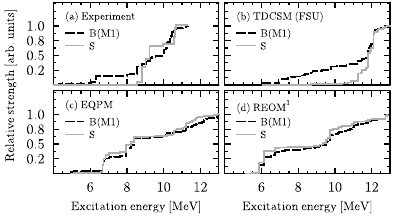}
    \caption{(a)-(d) Running sums of $B(M1)\uparrow$ and spectroscopic strengths $S$ normalized to unity. For the experimental $B(M1)$ strengths, the upper limit of the running sum is shown.}
    \label{fig:03}
\end{figure}

In the hope to understand these surprising results and being aware of the ongoing discussion of the basis dependence of spectroscopic factors, calculations were performed with three independent, state-of-the-art nuclear-structure models. These are the time-dependent continuum nuclear shell model (TDCSM) \cite{Vol09a, Vol14a} using the recently developed cross-shell FSU interaction \cite{Lub19a, Lub20a}, the energy density functional theory plus quasiparticle phonon model approach (EQPM) including up to three-phonon contributions \cite{Tso16a}, and the also recently developed relativistic equation of motion theory including the coupling of two-quasiparticles with up to two phonons (REOM$^3$) \cite{Lit19a, Lit22a, Lit23a}. The results for the $1^+$ states are shown in Fig.\,\ref{fig:02}. We note that the TDCSM calculations with the FSU interaction provide the same results as the conventional shell model approach. The TDCSM and REOM$^3$ predictions for the $B(M1)$ strengths were quenched using a typical factor of $q^2=(0.75)^2$ \cite{Lan04a}. For the EQPM, as is common practice \cite{SolV, Vdo79a}, the isovector spin-dipole coupling constant was fitted to reproduce the experimental $M1$ strength centroid of 9\,MeV. An effective gyromagnetic spin factor $g_s = 0.75$ was used. The running sums of the $M1$ strengths are shown in Fig.\,\ref{fig:02}\,(a). For all models, the summed strengths agree reasonably with data [see Figs.\,\ref{fig:02}\,(b)-(d)]. Even though shifted by about $+1.5$\,MeV relative to the data, the TDCSM predictions for the running sum agree best with the experimental data. The EQPM and REOM$^3$ also provide good descriptions of the running sum, and for the fragmentation of the $M1$ strengths into two major groups. In general, the details of the strength fragmentation are quite different between the models [see Figs.\,\ref{fig:02}\,(b)-(d)]. Looking at the spectroscopic factors, $S$, we can understand why that happens. As can be seen in Fig.\,\ref{fig:02}\,(e)-(g), all models expect states with larger neutron $(1f_{7/2})^{-1}(1f_{5/2})^{+1}$ spectroscopic factors at very different energies. They appear too high for the TDCSM, partly explaining why the $M1$ strength is shifted by $+1.5$\,MeV, and too low for the EQPM and REOM$^3$. For the EQPM and REOM$^3$, this also explains why there is a larger fraction of the $B(M1)$ strength at lower energies. Note that different from the TDCSM and EQPM, less spectroscopic strength is collected up to 13\,MeV in the REOM$^3$, i.e., only 68\,$\%$. This also leaves room for additional associated $B(M1)$ strength beyond 13\,MeV. Even though details are different, all three models expect that the largest $B(M1)$ values coincide with the largest $(1f_{7/2})^{-1}(1f_{5/2})^{+1}$ spectroscopic factors, $S$ [see Figs.\,\ref{fig:02}\,(b)-(g)]. This agrees with the general expectation that this configuration generates the bulk of the spin-flip $M1$ strength in \nuc{50}{Ti}. For the EQPM and REOM$^3$, the correlation becomes very clear when inspecting the relative $B(M1)$ and spectroscopic strengths in Fig.\,\ref{fig:03}. For the TDCSM, the discussion is more nuanced since, as in experiment, the $B(M1)$ strength suddenly increases when the $(1f_{7/2})^{-1}(1f_{5/2})^{+1}$ spectroscopic strength gets picked up [see Fig.\,\ref{fig:03}\,(b)]. Between 11\,MeV and 13\,MeV, 64\,$\%$ of the total $B(M1)$ strengths gets collected by states with larger $(1f_{7/2})^{-1}(1f_{5/2})^{+1}$ spectroscopic factors. The experimental upper limit for that fraction is 59\,$\%$. This number corresponds to the fraction this configuration would exhaust for the lower limit of the $B(M1)$ strength though. For the upper limit, it is only 19\,$\%$. Most importantly though, we want to make clear that the theoretically predicted correlation between larger spectroscopic factors and large $B(M1)$ strengths is in conflict with the experimental results (see Fig.\,\ref{fig:01} and pertaining discussion). In this letter, we showed for the first time that the states with larger neutron $(1f_{7/2})^{-1}(1f_{5/2})^{+1}$ spectroscopic factors do not correspond to the ones with the largest $B(M1)$ strengths.

For completeness, we comment on the structure of the 6.73-MeV, $1^+$ state predicted by the TDCSM. The wavefunction of this state is indeed dominated with 54\,$\%$ by the $(\pi:1f_{7/2})^2_{J^{\pi}=2^+_{\pi}}\otimes[(\nu:1f_{7/2})^{-1}(\nu:2p_{3/2})^{+1}]_{J^{\pi}=2^+_{\nu}}$ configuration. This is the microscopic structure of the expected orbital excitation mentioned in the introduction. In agreement with the experimental data discussed for the 6.307-MeV state, the ratio $R = B(\sigma)/B(M1)$ is 15\,$\%$ ($R_{exp} = 14(3)$\,$\%$) and the state has no significant $(1f_{7/2})^{-1}(1f_{5/2})^{+1}$ spectroscopic factor. Though it is close, the predicted $B(M1) = 0.59$\,$\mu_N^2$ value is smaller than the measured one. We do not claim a one-to-one correspondence. However, as shown by this example, orbital strength is not negligible and multiple observables needed to be interrogated for the comparison.

{\it Conclusions.} Two new experiments, $\nuc{50}{Ti}(\gamma,\gamma')$ and $\nuc{49}{Ti}(d,p)\nuc{50}{Ti}$, were performed to study the microscopic structure of $1^+$ states and the origin of $B(M1)$ strengths in the $N=28$ isotone \nuc{50}{Ti}. Specifically, the new data were compared to available $(e,e')$ and $(p,p')$ data to assess the role of the neutron $(1f_{7/2})^{-1}(1f_{5/2})^{+1}$ 1p-1h configuration in generating the spin-flip $M1$ strength. Using the information obtained with the complementary probes, this letter showed that $1^+$ states with larger $(1f_{7/2})^{-1}(1f_{5/2})^{+1}$ spectroscopic factors do not correspond to the ones with the largest $B(M1)$ strengths. This finding is in conflict with predictions made by three state-of-the-art nuclear structure models and disagrees with expectations that this configuration generates the bulk of spin-flip $M1$ strength in the $fp$ shell. Proton spin-flip contributions are expected at higher energies. It needs to be stressed that all models agree with the experimentally measured $M1$ strengths when typical quenching factors are employed. Nevertheless, the obvious disagreement for the detailed structure of the states begs the question whether it could contribute to the quenching of the $M1$ strength. Whatever causes the redistribution of the strength to other configurations seems to be currently missing in the models. The TDCSM includes configuration mixing and possible ground-state correlations within the full $fp$ shell. Thus, one possibility could be additional contributions from the unoccupied $1g$ and $3s2d$ shells, or even contributions from the lower $2s1d$ shell as shown to be relevant for \nuc{40}{Ca} \cite{Gro79a}. These configurations are not part of the current TDCSM configuration space. Contributions from the lower $sd$ shell would open up the possibility for $(1d_{5/2})^{-1}(1d_{3/2})^{+1}$ spin-flip excitations, which could not be probed in $(d,p)$, and for more orbital $M1$ strength. It is also important to note that the EQPM and REOM$^3$ cannot account for ground-state correlations at the moment. We cannot judge whether higher-order two-body currents provide a possible resolution \cite{Bra25a}. Our findings question, however, the theoretical analysis of the structure of $1^+$ states based on a comparison to their experimentally measured $B(M1)$ strength alone. Going forward, it will be necessary to consider all $1^+$ states and associated experimental observables to guarantee a meaningful comparison as the strength is clearly fragmented among several states.

{\it Acknowledgments.} This work was supported by the U.S. National Science Foundation under Grant Nos. PHY-2012522 (FSU), PHY-2412808 (FSU), PHY-2209376 (WMU), and PHY-2515056 (WMU), and by the U.S. Department of Energy, Office of Science, Office of Nuclear Physics under Award Nos. DE-SC0009883 (FSU) and DE-FG02-97ER41033 (Duke U.). This work was also supported by the Deutsche Forschungsgemeinschaft (DFG, German Research Foundation) under project-ID 499256822 - GRK 2891 ``Nuclear Photonics'', and by project ELI-RO/DFG/2023\_001 ARNPhot funded by the Institute of Atomic Physics, Romania. Parts of this work were carried out under the contract PN 23.21.01.06 sponsored by the Romanian Ministry of Research, Innovation and Digitalization and partially supported by ELI-RO-RDI-2024-AMAP of the Romanian Government. A target provided by the Center for Accelerator Target Science at Argonne National Laboratory was used in this work. The authors thank Simela Aslanidou, Jakob Beller, Vera Derya, Ralph Kern, Andreas Krugmann, Bastian L{\"o}her, Laura Mertes, Thomas M{\"o}ller, Philipp Ries, Christopher Romig, Johannes Wiederhold, Julius Wilhelmy, and Markus Zweidinger for their support during data taking. M.S. expresses his gratitude to K.W. Kemper, J. Piekarewicz, P. von Neumann-Cosel, and A. Zilges for their critical input.

{\it Data availability.} The data are not publicly available. The data are available from the authors upon reasonable request.

\bibliography{references}

\end{document}